\documentclass[letterpaper,onecolumn,accepted=2021-04-06]{quantumarticle}
\pdfoutput=1
\usepackage[utf8]{inputenc}
\usepackage{amsthm,amsfonts,framed}

\DeclareMathSizes{10}{9}{7}{5}

\usepackage{physics, graphicx, hyperref}

\theoremstyle{plain}
\newtheorem*{thm}{Theorem} 

\begin{document}
\title{Local classical MAX-CUT algorithm outperforms $p=2$ QAOA on high-girth regular graphs}
\author{Kunal Marwaha}
\orcid{0000-0001-9084-6971}
\email{marwahaha@berkeley.edu}
\affiliation{Berkeley Center for Quantum Information and Computation, University of California, Berkeley, California 94720, USA}

\begin{abstract}
The $p$-stage Quantum Approximate Optimization Algorithm (QAOA$_p$) is a promising approach for combinatorial optimization on noisy intermediate-scale quantum (NISQ) devices, but its theoretical behavior is not well understood beyond $p=1$. We analyze QAOA$_2$ for the \emph{maximum cut problem} (MAX-CUT), deriving a graph-size-independent expression for the expected cut fraction on any $D$-regular graph of girth $> 5$ (i.e. without triangles, squares, or pentagons).

We show that for all degrees $D \ge 2$ and every $D$-regular graph $G$ of girth $> 5$, QAOA$_2$ has a larger expected cut fraction than QAOA$_1$ on $G$. However, we also show that there exists a $2$-local randomized \emph{classical} algorithm $A$ such that $A$ has a larger expected cut fraction than QAOA$_2$ on all $G$. This supports our conjecture that for every constant $p$, there exists a local classical MAX-CUT algorithm that performs as well as QAOA$_p$ on all graphs.
\end{abstract}

\maketitle

\section{Introduction}

The $p$-stage Quantum Approximate Optimization Algorithm (QAOA$_p$) \cite{farhi2014quantum} is a protocol to use near-term quantum computers for combinatorial optimization \cite{Preskill_2018}. The number $p$ is called the algorithm's \emph{depth}, with small values of $p$ realizable on current devices \cite{arute2020quantum} \cite{Zhou_2020}, often called noisy intermediate-scale quantum (NISQ) devices. The performance of the algorithm is difficult to analyze, even for small values of $p$ and in restricted settings \cite{farhi2014quantum} \cite{wurtz2020bounds}. In this work we analyze QAOA$_2$ for the \emph{maximum cut problem} (MAX-CUT) and compare it with \emph{local classical algorithms}. There is a classical algorithm (the Goemans-Williamson semidefinite program \cite{goemans1995improved}) that, under certain computational complexity assumptions, achieves the optimal worst-case approximation ratio for MAX-CUT among all polynomial-time classical and quantum algorithms \cite{Khot02}.\footnote{Specifically, this holds assuming Khot's \emph{Unique Games Conjecture} and the widely-believed conjecture that ${\mathbf{NP} \not\subseteq \mathbf{BQP}}$.} However, this is a worst-case ratio; there may be inputs where QAOA$_p$ outperforms the Goemans-Williamson program. Moreover, the Goemans-Williamson program is a \emph{global} algorithm, for which the assignment of each vertex depends on the entire graph; by contrast, QAOA$_p$ is a $p$\emph{-local} algorithm, where the assignment depends only on the vertex's radius $p$ neighborhood. Does there exist a local classical algorithm that performs as well as QAOA$_p$ on every graph? In this work we answer this question positively for $p = 2$ on regular graphs of girth above 5.

\subsection{Related work}

QAOA is thought to be competitive with classical approximation algorithms. Initially, QAOA$_1$ was the best known approximate algorithm for MAX-3-LIN-2 \cite{farhi2015quantum}, although an improved classical algorithm was quickly found \cite{barakmaxklin2}. For MAX-CUT, QAOA$_1$ guarantees an expected cut fraction of $> \frac{1}{2} + \frac{0.3032}{\sqrt{D}}$ on triangle-free $D$-regular graphs \cite{Wang2018}, outperforming the lower bound of the best known local classical algorithm, the \emph{threshold algorithm} \cite{hrss}. However, a direct calculation by Hastings \cite{hastings2019classical} shows that this lower bound is not tight: The threshold algorithm with optimal parameter value outperforms QAOA$_1$ on triangle-free $D$-regular graphs for all but $4$ choices of $2 \le D < 1000$, and likely for all larger $D$. Hastings also introduces a family of local classical optimization algorithms, identifying algorithms from this family that outperform QAOA$_1$ for each of the remaining $4$ choices of $D$.
\newline

QAOA has very few proven results about its performance, and QAOA$_p$'s optimal parameters are only known for small $p$ or for severely restricted problems.  \cite{Wang2018} derives a graph-size-independent expression of the expected cut fraction for QAOA$_1$ on any graph, determining the maximum for any triangle-free $D$-regular graph. \cite{Wang2018} and \cite{szegedy2019qaoa} use separate approaches to derive the expected and maximum expected cut fraction for QAOA$_2$ on any $2$-regular graph. \cite{wurtz2020bounds} shows that QAOA$_2$ has a maximum expected cut fraction of approximately $0.7559$ on $3$-regular graphs. For larger $p$, the creators of QAOA have proposed a hybrid algorithm to find optimal parameters, alternating between the quantum circuit and a classical parameter optimizer; unfortunately, this approach involves non-convex classical optimization, making its runtime difficult to analyze \cite{McClean_2016}.

\subsection{Our results}
This work studies QAOA$_2$ and local classical algorithms for MAX-CUT on $D$-regular graphs of girth $> 5$. We derive a graph-size-independent expression for the expected cut fraction of QAOA$_2$ on these graphs. We numerically optimize this expression over QAOA's input parameters to find the maximum expected cut fraction for each $D < 500$. We then generalize the $1$-local threshold algorithm with one parameter $\tau$ to the $n$\emph{-step threshold algorithm}, an $n$-local algorithm with $n$ parameters $(\tau_1, \cdots, \tau_n)$. We derive the performance of the $2$-step threshold algorithm on these graphs as a function of $(\tau_1, \tau_2)$. When $\tau_1 = \tau_2$, a direct calculation shows that the optimal $2$-step threshold algorithm outperforms QAOA$_2$ for all $41 < D < 500$, and likely for all larger $D$ based on asymptotic behavior. Another direct calculation on $D < 50$ shows that the optimal $2$-step threshold algorithm outperforms QAOA$_2$ for all $5 < D < 50$. We identify the $2$-step threshold algorithm with the family of 2-local classical algorithms in \cite{hastings2019classical}, specifying instances that outperform QAOA$_2$ for the remaining 4 choices. This shows that for all $D \ge 2$, there is a $2$-local classical MAX-CUT algorithm that outperforms QAOA$_2$ on all $D$-regular graphs of girth above 5.
\newline

Why the restriction on girth? This condition massively simplifies the analysis of local algorithms. Consider an $\ell$-local algorithm, where a vertex's assignment depends on its radius $\ell$ neighborhood. When girth $> 2\ell+1$, the neighborhood looks like a tree. Previous analyses of QAOA$_1$ and the 1-step threshold algorithm use this property when studying triangle-free graphs (i.e. girth $> 3$). Similarly, our analysis of QAOA$_2$ and the 2-step threshold algorithm considers graphs of girth above $5$ (no triangles, squares, or pentagons). Our analysis could be extended beyond $\ell=2$; we suspect that the $p$-step threshold algorithm performs as well as QAOA$_p$ on every graph of girth $> 2p+1$.

\section{QAOA$_2$ performance on MAX-CUT}
\label{sec:qaoa2formula}
Given a combinatorial problem on a graph $G(V, E)$, QAOA is an algorithm on $|V|$ qubits, where each vertex in $G$ corresponds to a qubit. QAOA involves two Hamiltonians: the mixing Hamiltonian $J = \sum_{i\in V} \sigma_i^x$ and the problem Hamiltonian $C$, where each eigenstate of $C$ encodes a possible solution scored by its eigenvalue. The algorithm approximates adiabatic evolution from the maximal eigenstate of $J$ to the maximal eigenstate of $C$, the optimal solution. Precisely, QAOA$_p$ takes $2p$ parameters $(\gamma_1, \beta_1, \cdots, \gamma_p, \beta_p)$ and evolves the initial state $\rho_0 = \bigotimes_{i\in V} (I/2 + \sigma_i^x/2)$ with a unitary $U =  e^{-i\beta_p J} e^{-i\gamma_p C} \cdots  e^{-i\beta_1 J} e^{-i\gamma_1 C}$.  The state is then measured in the eigenbasis of $C$, which gives an approximate solution. More information on QAOA can be found in \cite{farhi2014quantum} and \cite{Wang2018}.
\newline

For MAX-CUT, let $C =  \sum_{(u, v) \in E} C_{uv}$ where $C_{uv} =0.5(1- \sigma_u^z \sigma_v^z)$. The maximal eigenvalue of $C$ is the maximum number of edges that can be cut on the graph $G$; an edge is ``cut" if the qubits corresponding to its vertices do not agree when measured in the $z$-basis. The expected number of edges cut by QAOA is $\Tr[CU\rho_0 U^\dagger] = \sum_{(u, v) \in E} \Tr[\rho_0 U^\dagger C_{uv} U ]$.
\newline

Let $f_{p,uv}$ represent the chance that QAOA$_p$ cuts the edge $(u,v)$ of graph $G$. For $p=2$:
$$
f_{2,uv}(\gamma_1, \beta_1, \gamma_2, \beta_2)
= \Tr[\rho_0 e^{i\gamma_1 C} e^{i\beta_1 J} e^{i\gamma_2 C} e^{i\beta_2 J} 
C_{uv}
e^{-i\beta_2 J} e^{-i\gamma_2 C} e^{-i\beta_1 J} e^{-i\gamma_1 C} ]
$$

This is hard to evaluate in general because of the dependence on $C$ and $J$. However, $f_{p,uv}$ only involves terms within the radius $p$ subgraph from vertices $u$ and $v$. If this subgraph is identical for every edge, $f_{p,uv} = f_p$ represents the expected cut fraction of QAOA$_p$ on the graph. For example, this happens for graphs of girth $> 2p+1$.

\begin{thm} Consider any $D$-regular graph of girth $> 5$. Then:
$$
f_{2}(\gamma_1, \beta_1, \gamma_2, \beta_2) = \frac{1}{2} +  c^2 rt y^{D-1}z - \frac{cs}{2} \alpha(\gamma_1, \beta_1, \gamma_2) - \frac{s^2 tz}{4} \kappa(\gamma_1, \beta_1, \gamma_2)
$$
where
\begin{align*}
c&=\cos(2 \beta_2) & m&= \cos(\gamma_2) & r&=\cos(2 \beta_1) & y&=\cos(\gamma_1)
\\
s&=\sin(2 \beta_2) & n&=\sin(\gamma_2) & t&=\sin(2 \beta_1)  & z&=\sin(\gamma_1)
\end{align*}
and
\begin{align*}
    \kappa(\gamma_1, \beta_1, \gamma_2) &= \big((1+r)(my - nrz)^{D-1} - (1-r)(my + nrz)^{D-1}\big)
    \\
    &\times \big( (m + inty^{D-1}z)^{D-1} + (m - inty^{D-1}z)^{D-1} \big)
\end{align*}
and
\begin{multline*}
\alpha(\gamma_1, \beta_1, \gamma_2) = (1+r)(-mrz - ny)(my - nrz)^{D-1} +(1-r) (mrz - ny)(my + nrz)^{D-1}
\\ + t \big ( (mty^{D-1}z + in)(m + inty^{D-1}z)^{D-1} + (mty^{D-1}z - in)(m - inty^{D-1}z)^{D-1} \big )
\end{multline*}

\end{thm}

This formula notably does not depend on graph size $|V|$ or the Hilbert space; it does not require evaluating a trace. See Appendix \ref{sec:proofqaoa2formula} for a proof. We also validate this formula by simplifying to known expressions; see Appendix \ref{sec:correctness_qaoa2formula} for details.
\newline

Using SciPy \cite{2020SciPy-NMeth}, we numerically optimize the above formula for each $2 \le D < 500$. Figure \ref{fig:all_results} shows the optimized performance of QAOA$_2$ for $D < 500$. In all cases, QAOA$_2$ outperforms both optimal QAOA$_1$ and the 1-local classical algorithms described in Section III of \cite{hastings2019classical}. See Appendix \ref{tx} for the numerical values at small $D$. The values at $D=2$ and $D=3$ reproduce known results by \cite{Wang2018}, \cite{szegedy2019qaoa}, and \cite{wurtz2020bounds}. For $D=3$, \cite{wurtz2020bounds} finds the full set of optimal input parameters; we verify that our input parameters are in this set.

\section{Local classical algorithms}
Below is a simplified description of the 1-step threshold algorithm, first presented in \cite{hrss}.
\begin{quote}
Consider a graph and threshold $\tau$. Randomly assign each vertex a spin $+1$ or $-1$ with equal probability. Then, consider any vertices with the same spin as $\ge \tau$ of its neighbors. Flip the spin of those vertices. Cut the graph into ``spin $+1$" and ``spin $-1$" partitions.
\end{quote}

We propose a $n$-step variation of the threshold algorithm, where the ``Consider" and ``Flip" commands are run $n$ times before making a cut, using threshold $\tau_i$ for $i \in \{1, \ldots, n\}$.
\begin{quote}
Consider a graph and thresholds $\tau_1, \cdots, \tau_n$. Randomly assign each vertex a spin $+1$ or $-1$ with equal probability. Then, for $i = 1 \ldots n$: consider any vertices with the same spin as $\ge \tau_i$ of its neighbors, and flip the spin of those vertices. Cut the graph into ``spin $+1$" and ``spin $-1$" partitions.
\end{quote}
The independence condition that simplifies triangle-free analysis of 1-local algorithms gets more restrictive with many steps. A similar condition on the $n$-step threshold algorithm requires the graph to have girth $> 2n+1$, i.e. no cycles of length less than $2n + 2$. However, unlike QAOA$_n$, the $n$-step threshold algorithm has no guarantee of achieving the maximum cut fraction as $n \to \infty$.
\newline

\begin{figure}[ht]
    \centering
    \includegraphics[width=\linewidth]{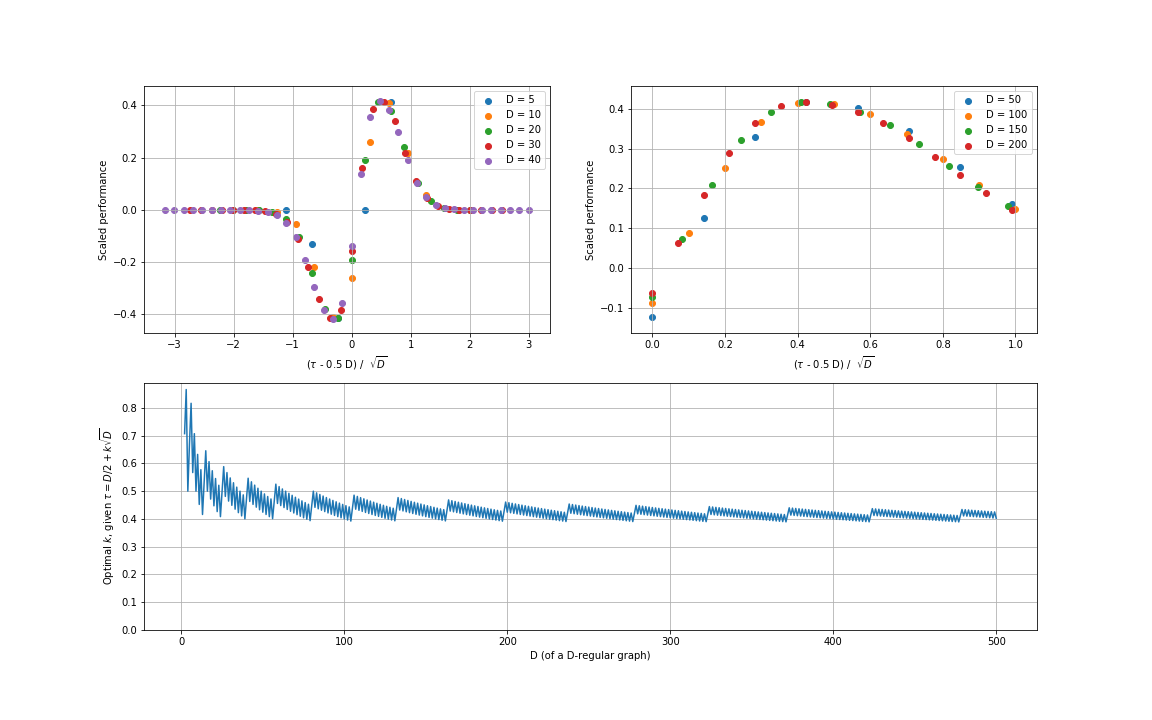}
    \caption{Top: Scaled performance of the 2-step threshold algorithm compared to threshold $\tau_1= \tau_2$. The scaled performance is $b$ such that the cut fraction is $1/2 + bD^{-0.5}$. The left plot matches results from the 1-step threshold algorithm, as shown in Figure 6 of \cite{hrss}. The right plot uses larger values of $D$ with a comparatively smaller region of thresholds. Note that the optimal threshold is near $D/2 + 0.4\sqrt{D}$ for any $D$-regular graph. This informed our search space to locate the optimal threshold at larger $D$.
    Bottom: Optimal threshold for the 2-step threshold algorithm where $\tau_1 = \tau_2$. Notice the convergence on a value close to $D/2 + 0.4\sqrt{D}$. We use a limited search space at larger $D$: within ($D/2 + k_{min}\sqrt{D}$, $D/2 + k_{max}\sqrt{D}$), where $(k_{min},  k_{max}) = (0.3, 0.6)$ for $60 < D \le 150$, and $(k_{min},  k_{max}) = (0.35, 0.52)$ for $D > 150$. The optimal $k$ at large $D$ is quite close to that for the 1-step threshold algorithm, as presented in Figures 7-8 of \cite{hrss}.}
    \label{fig:best_threshold}
\end{figure}

We derive an expression for the performance of the 2-step threshold algorithm on graphs of girth $> 5$ as a function of $\tau_1, \tau_2$. The derivation is in Appendix \ref{sec:threshold2_analysis}. We directly calculate the maximum value across all thresholds for small values of $D$, first assuming $\tau_1 = \tau_2$. The optimal threshold approximately matches the value given in \cite{hrss}: $\tau = D/2 + k \sqrt{D}$, where $k \approx 0.4$. For intermediate values of $D$ up to 500, we limit our search of optimal $\tau_1, \tau_2$. See Figure \ref{fig:best_threshold} for details. 
\newline

\begin{figure}[ht]
    \centering
    \includegraphics[width=\linewidth]{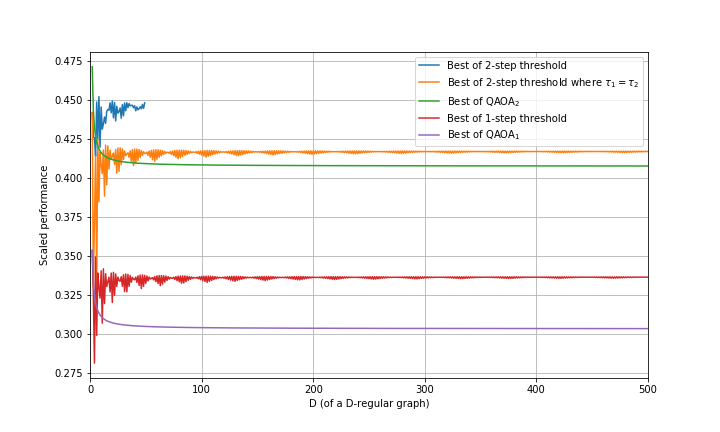}
    \caption{Performance of approximate MAX-CUT algorithms on $D$-regular graphs of girth $>5$. This plot reproduces and extends Figure 3 of \cite{hastings2019classical}. The scaled performance $b$ is such that the cut fraction is $1/2 + bD^{-0.5}$. For QAOA$_2$, the plot is smooth and similar in shape to QAOA$_1$, while the 2-step threshold algorithm oscillates like the 1-step variant. At large degree, $b \approx 0.417$ for the 2-step threshold algorithm, which outperforms $b \approx 0.407$ for QAOA$_2$, itself an improvement over both $b \approx 0.33$ for the 1-step threshold algorithm and $b \approx 0.303$ for QAOA$_1$. Judging by the convergence at larger $D$, we expect the 2-step classical algorithm to outperform QAOA$_2$ for all $D \ge 500$.}
    \label{fig:all_results}
\end{figure}

When the thresholds match, the performance of this algorithm stabilizes at $0.5 + b/\sqrt{D}$, where $b \approx 0.417$. This outperforms QAOA$_2$ for all $41 < D < 500$. Considering Figure \ref{fig:all_results} in a similar spirit to Figure 3 of \cite{hastings2019classical}, we expect the 2-step threshold algorithm to outperform QAOA$_2$ for all $D \ge 500$. Note the oscillations that decrease in value for large $D$, similar to those of the 1-step threshold algorithm. As \cite{hastings2019classical} suggests, this likely happens from optimizing a discrete parameter $\tau$ instead of a continuous one. We also consider thresholds where $\tau_1$ may not equal $\tau_2$ in general for all $2 \le D < 50$; when $D > 5$, there are choices of $\tau_1, \tau_2$ that outperform QAOA$_2$. See Figure \ref{fig:all_results} for a comparison.
\newline 

For $D \in [2, 3, 4, 5]$, we draw inspiration from modifications made to the 1-step threshold algorithm in Section III of \cite{hastings2019classical}. Using the local algorithm description from  \cite{hastings2019classical}, we consider a 2-step linear algorithm with entries of $\vec{v_0}$ chosen randomly from $[-1, 1]$. Because there are a finite number of initial values, we can exactly calculate the expected performance on a local subgraph. We search for values of $(c_0, c_1)$ that outperform QAOA$_2$. This was successful for $D \in [3, 4, 5]$, but did not work for $D=2$. In this case, we also searched over choices of initial values to find parameters that outperform QAOA$_2$. In particular, given initial values $[-0.49, -0.45, 0.01, 0.03, 0.29, 0.85]$, a $D=2$ local classical algorithm has expected performance $0.3343$ over random assignment, whereas QAOA$_2$ has at most $1/3$ over random assignment \cite{mbeng2019quantum}. Many choices of initial values give a $D=2$ classical algorithm with maximum expected performance near this value; we are unsure why this is the case.
\newline

Thus, there exists a local classical MAX-CUT algorithm that, on average, finds a larger cut than QAOA$_2$ on every $D$-regular graph of girth above 5 when $D < 500$, and likely for all $D$. The expected performance values at small $D$ are reproduced in Appendix \ref{tx}. An interactive notebook \cite{jupyter} with all relevant code and figures is \href{https://nbviewer.jupyter.org/github/marwahaha/qaoa-local-competitors/blob/master/2-step-comparison.ipynb}{available online}.

\section{Discussion}
QAOA$_p$'s optimal performance is guaranteed to increase with increasing $p$. At $p=2$, this value was previously known only in a few cases \cite{Wang2018} \cite{wurtz2020bounds}. Our approach is to directly calculate a graph-size-independent formula and optimize over parameters, which requires careful accounting of all terms. Hypothetically, Appendix \ref{sec:proofqaoa2formula} can be extended to remove the high-girth and regularity constraints. Without the regularity constraint, a graph cannot be described with a single parameter $D$; without the high-girth constraint, there may be implicit combinatorial sums as in Appendix A of \cite{Wang2018}.
\newline

We wonder if QAOA$_p$'s optimal performance on high-girth graphs is more connected to ``locality" than ``quantumness". The high-girth condition makes the graph look like a tree to any local algorithm. Given any constant $p$, we expect that a graph-size-independent expression can be efficiently calculated for both QAOA$_p$ and the $p$-step threshold algorithm on regular graphs of girth $> 2p+1$; we suspect the $p$-local classical algorithm, on average, will find a larger cut than QAOA$_p$ on every such graph.
\newline

Importantly, we compare QAOA$_2$ with \emph{local} classical MAX-CUT algorithms. This constraint may be useful in distributed settings, where computing nodes do not have access to the entire graph and communication between computing nodes is limited. Even so, this work suggests that QAOA$_p$ is outperformed by a local classical algorithm of similar depth. We conjecture that for every constant $p$, there exists a local classical algorithm that, on average, finds a cut at least as large as QAOA$_p$ on every possible graph.
\newline

We only study $D$-regular graphs. If the graph has instead maximum degree $D$, one can still use this analysis by instituting ``phantom vertices", where random data is generated for each missing vertex. We first learned of this idea from \cite{shearer1992} but do not think it is optimal. We also compare graphs only of girth above 5. If there are a small number of short cycles, we do not expect this result to change; however, threshold algorithms are not very effective on highly connected graphs, a topic we plan to study further.

\section*{Acknowledgements}
Thanks to Ryan Avery and Sangeeta Marwaha for helpful discussions. Thanks to Boaz Barak for detailed comments on the abstract and introduction. Thanks to Patrick Lutz, Bryan O'Gorman, and Akshay Ramachandran for reading a draft and providing suggestions. Thanks to Birgitta Whaley and everyone at BQIC for listening to a talk of preliminary results. 

\bibliographystyle{halpha-abbrv-mod}
\bibliography{research}

\clearpage
\appendix
\pagebreak

\section{Listing of performance values at small $D$}
\label{tx}

\begin{table}[!htb]
\resizebox{\textwidth}{!}{
\begin{tabular}{l|l|l|l|l|l|l|l}
D & QAOA$_1$ & Threshold$_1$ ($\tau$) & Modified Threshold$_1$ & QAOA$_2$ & Threshold$_2$ ($\tau_1$, $\tau_2$) & Modified Threshold$_2$ ($c_0$, $c_1$) \\
\hline
2 & 0.2500 & 0.2500 (2) & \: --- & 0.3333 & 0.3125 (2, 2) & 0.3343 (17.96, 0.56) \\
3 & 0.1925 & 0.1875 (3) & 0.1980 & 0.2559 & 0.2461 (2, 3) & 0.2703 (7.22, 0.58) \\
4 & 0.1624 & 0.1406 (3) & 0.2161 & 0.1693 & 0.2128 (3, 4) & 0.2309 (0.38, 11.20) \\
5 & 0.1431 & 0.1562 (4) & \: --- & 0.1907 & 0.1851 (4, 4) & 0.2045 (5.44, 0.33) \\
6 & 0.1294 & 0.1221 (5) & 0.1302 & 0.1726 & 0.1832 (4, 5) & \: --- \\
7 & 0.1190 & 0.1282 (5) & \: --- & 0.1589 & 0.1607 (5, 5) & \: --- \\
8 & 0.1108 & 0.1166 (6) & \: --- & 0.1480 & 0.1599 (5, 6) & \: --- \\
9 & 0.1040 & 0.1077 (6) & \: --- & 0.1391 & 0.1399 (5, 7) & \: --- \\
10 & 0.0984 & 0.1077 (7) & \: --- & 0.1317 & 0.1409 (6, 7) & \: --- \\
11 & 0.0936 & 0.0925 (7) & 0.0986 & 0.1253 & 0.1301 (7, 8) & \: --- \\
12 & 0.0894 & 0.0987 (8) & \: --- & 0.1197 & 0.1254 (7, 8) & \: --- \\
13 & 0.0858 & 0.0886 (9) & \: --- & 0.1149 & 0.1218 (8, 9) & \: --- \\
14 & 0.0825 & 0.0905 (9) & \: --- & 0.1106 & 0.1163 (8, 10) & \: --- \\
15 & 0.0796 & 0.0853 (10) & \: --- & 0.1067 & 0.1143 (9, 10) & \: ---  \\
16 & 0.0770 & 0.0833 (10) & \: --- & 0.1032 & 0.1110 (9, 11) & \: --- \\
17 & 0.0747 & 0.0816 (11) & \: --- & 0.1001 & 0.1076 (10, 11) & \: --- \\
18 & 0.0725 & 0.0771 (11) & \: --- & 0.0972 & 0.1056 (10, 12) &  \: --- \\
19 & 0.0705 & 0.0778 (12) & \: --- & 0.0945 & 0.1014 (11, 12) & \: ---
\end{tabular}}
\caption{Improvement over random assignment for MAX-CUT on regular graphs of girth above 5.  The degree $D$ is shown in the first column. ``Threshold$_1$ ($\tau$)", ``Modified Threshold$_1$" and ``QAOA$_1$" are reproduced from Table I of \cite{hastings2019classical}. The QAOA$_2$ and 2-step threshold outperform all previously known values at each listed $D$, and all $D < 500$ as shown numerically. ``Threshold$_2$" refers to the best 2-step threshold algorithm with $\tau_1$ and $\tau_2$ possibly unequal. It does not outperform QAOA$_2$ when $D\in [2,3,4,5]$, but a related 2-step local classical algorithm using \cite{hastings2019classical}'s framework does.}
\end{table}

\clearpage

\section{Proof of the $p=2$ QAOA formula}
\label{sec:proofqaoa2formula}
This analysis follows the approach given in Appendix A of \cite{Wang2018}. Assume the graph is $D$-regular, and has no triangles, squares, or pentagons (girth $> 5$). In addition to the variables defined in \ref{sec:qaoa2formula}, let $p = \cos(\gamma_1 / 2)$, $q = \sin(\gamma_1 / 2)$, $k = \cos(\beta_1)$, and $l = \sin(\beta_1)$. Then, by double-angle formulas, $r = k^2 - l^2$, $t = 2kl$, $y = p^2 - q^2$, and $z = 2pq$. We start with the expression for $f_2(\gamma_1, \beta_1, \gamma_2, \beta_2)$, simplifying $C_{uv}$ as in Appendix A of \cite{Wang2018}:
$$
f_2(\gamma_1, \beta_1, \gamma_2, \beta_2)
= \frac{1}{2} - \frac{1}{2} \Tr[\rho_0 e^{i\gamma_1 C} e^{i\beta_1 J} e^{i\gamma_2 C}
\Big(c^2 \sigma_u^z \sigma_v^z + cs(\sigma_u^y \sigma_v^z + \sigma_u^z \sigma_v^y) + s^2 \sigma_u^y \sigma_v^y \Big)
e^{-i\gamma_2 C} e^{-i\beta_1 J} e^{-i\gamma_1 C} ]
$$

We compute each of the four terms inside the trace separately, labeled so that $f_2 = 0.5 - 0.5 (A + B_1 + B_2 + E)$.

\subsection{Calculating $A$}
Since $C$ commutes with $\sigma_u^z\sigma_v^z$, $f_1 = 0.5 - 0.5 A/c^2$. As the graph is $D$-regular and has no triangles:
$$
A/c^2 = -\sin(4\beta_1)\sin(\gamma_1)\cos^{D-1}(\gamma_1) = -2rt y^{D-1} z
$$

\subsection{Calculating $B_1$ and $B_2$}

\subsubsection{Evaluating exponentials}
Since the graph is identical from $u$ and $v$'s perspective, $B_1 = B_2$.
$$
B_1/cs = \Tr[\rho_0 e^{i\gamma_1 C} e^{i\beta_1 J} e^{i\gamma_2 C}
\sigma_u^y \sigma_v^z 
e^{-i\gamma_2 C} e^{-i\beta_1 J} e^{-i\gamma_1 C} ]
$$

The only terms of $C$ that do not commute with $\sigma_u^y \sigma_v^z$ involve nodes $w_1, ... w_{D-1}, v$ adjacent to $u$.
$$
B_1/cs = \Tr[\rho_0 e^{i\gamma_1 C} e^{i\beta_1 J}
\big( \prod_{i=1}^{D-1} e^{-i \gamma_2 \sigma_u^z \sigma_{w_i}^z} \big) 
e^{-i \gamma_2 \sigma_u^z \sigma_v^z}
\sigma_u^y \sigma_v^z 
 e^{-i\beta_1 J} e^{-i\gamma_1 C} ]
$$

The only terms of $J$ that do not commute with the inner expression reference the nodes $u, v, w_1, ..., w_{D-1}$.
$$
B_1/cs = \Tr[\rho_0 e^{i\gamma_1 C}
e^{i\beta_1 \sigma_u^x}
\big( \prod_{i=1}^{D-1} 
  e^{i\beta_1 \sigma_{w_i}^x} 
  e^{-i \gamma_2 \sigma_u^z \sigma_{w_i}^z}
  e^{-i\beta_1 \sigma_{w_i}^x} 
\big) 
\big(  e^{i\beta_1 \sigma_v^x} e^{-i \gamma_2 \sigma_u^z \sigma_v^z}  
  e^{i\beta_1 \sigma_v^x} \big)
e^{i\beta_1 \sigma_u^x} 
\sigma_u^y \sigma_v^z 
e^{-i\gamma_1 C} ]
$$

The terms in parentheses can be simplified.
$$
e^{i \beta_1 \sigma_{w_i}^x}  e^{-i \gamma_2 \sigma_u^z \sigma_{w_i}^z} e^{-i\beta_1 \sigma_{w_i}^x}
 = e^{i \beta_1 \sigma_{w_i}^x} (m - in \sigma_u^z \sigma_{w_i}^z) e^{-i \beta_1 \sigma_{w_i}^x}
 = m - in e^{2 i \beta_1 \sigma_{w_i}^x}\sigma_u^z \sigma_{w_i}^z
$$
$$
e^{i \beta_1 \sigma_{v}^x}  e^{-i \gamma_2 \sigma_u^z \sigma_{v}^z} e^{i\beta_1 \sigma_{v}^x}
 = e^{i \beta_1 \sigma_{v}^x} (m - in \sigma_u^z \sigma_{v}^z) e^{i \beta_1 \sigma_{v}^x}
 = e^{2 i \beta_1 \sigma_{v}^x}m - in \sigma_u^z \sigma_{v}^z
$$
$$
B_1/cs = \Tr[\rho_0 e^{i\gamma_1 C}
e^{i\beta_1 \sigma_u^x}
\big( \prod_{i=1}^{D-1} m - in e^{2 i \beta_1 \sigma_{w_i}^x}\sigma_u^z \sigma_{w_i}^z \big) 
\big(  e^{2 i \beta_1 \sigma_{v}^x}m - in \sigma_u^z \sigma_{v}^z \big)
e^{i\beta_1 \sigma_u^x} 
\sigma_u^y \sigma_v^z 
e^{-i\gamma_1 C} ]
$$

\subsubsection{Breaking apart into terms}

The only terms of $C$ that may not commute involve edges connected to a node in $S = \{u, v, w_1, ..., w_{D-1}\}$. Terms of $e^{i\gamma_1 C}$ that connect $j$ and $k$ have the form $p - iq \sigma_j^z \sigma_k^z$. Consider edges where both $j,k \in S$; we discuss the other edges in Appendix \ref{sub:b1crossover}.
\begin{multline*}
B_1/cs =  
\text{Tr}\Bigg[ \rho_0 
(p - iq \sigma_u^z \sigma_v^z)
\big( \prod_{i=1}^{D-1} (p - iq \sigma_u^z \sigma_{w_i}^z) \big)
\\ e^{i\beta_1 \sigma_u^x}
\big( \prod_{i=1}^{D-1} (m - in e^{2 i \beta_1 \sigma_{w_i}^x}\sigma_u^z \sigma_{w_i}^z) \big) 
\big(  e^{2 i \beta_1 \sigma_{v}^x}m - in \sigma_u^z \sigma_{v}^z \big)
e^{i\beta_1 \sigma_u^x} 
\\ (p - iq \sigma_u^z \sigma_v^z)
\big( \prod_{i=1}^{D-1} (p - iq \sigma_u^z \sigma_{w_i}^z) \big) \sigma_u^y \sigma_v^z
\Bigg]
\end{multline*}

We will expand the products carefully. Many terms do not contribute to trace. Only half include an even number of $\sigma_v^z$; the rest will not contribute to trace because of the definition of $\rho_0$. Note that each of these terms includes an odd number of $\sigma_u^z$, correctly cancelling out the rightmost $\sigma_u^y$ in the trace. Consider the terms that include $\sigma_v^z$:
$$
(p - iq \sigma_u^z \sigma_v^z) \cdots (e^{2 i \beta_1 \sigma_{v}^x}m - in \sigma_u^z \sigma_{v}^z)
\cdots (p - iq \sigma_u^z \sigma_v^z) \sigma_v^z
$$

\begin{enumerate}
    \item $(-iq \sigma_u^z \sigma_v^z, e^{2 i \beta_1 \sigma_{v}^x}m , p\sigma_v^z) \to V_1 =  -impqe^{-2i\beta_1 \sigma_{v}^x}$
    \item  $(p, - in \sigma_u^z \sigma_{v}^z , p\sigma_v^z) \to V_2 =  -inp^2$
    \item $(p, e^{2 i \beta_1 \sigma_{v}^x}m, -iq \sigma_u^z \sigma_v^z\sigma_v^z) \to V_3 = -impqe^{2i\beta_1 \sigma_{v}^x}$
    \item $(-iq \sigma_u^z \sigma_v^z, - in \sigma_u^z \sigma_{v}^z , -iq \sigma_u^z \sigma_v^z\sigma_v^z) \to V_4 = inq^2$
\end{enumerate}

Similarly, of the expressions that include $\sigma_{w_i}^z$, only half include an even number. Only these corresponding terms can contribute to trace. Note that each of these terms includes an even number of $\sigma_u^z$, which will cancel out. Sometimes the exponential sign flips when commuting with $\sigma^z$.
$$
(p - iq \sigma_u^z \sigma_{w_i}^z)
 \cdots (m - in e^{2 i \beta_1 \sigma_{w_i}^x}\sigma_u^z \sigma_{w_i}^z)
\cdots (p - iq \sigma_u^z \sigma_{w_i}^z)
$$
\begin{enumerate}
    \item $(p, m, p) \to W_1 = mp^2$
    \item  $(p, - in e^{2 i \beta_1 \sigma_{w_i}^x}\sigma_u^z \sigma_{w_i}^z, - iq \sigma_u^z \sigma_{w_i}^z) \to W_2 = -npqe^{2 i \beta_1 \sigma_{w_i}^x}$
    \item $(- iq \sigma_u^z \sigma_{w_i}^z, m, - iq \sigma_u^z \sigma_{w_i}^z) \to W_3 = -mq^2$
    \item $( - iq \sigma_u^z \sigma_{w_i}^z, - in e^{2 i \beta_1 \sigma_{w_i}^x}\sigma_u^z \sigma_{w_i}^z, p) \to W_4 = -npq e^{-2 i \beta_1 \sigma_{w_i}^x}$
\end{enumerate}

The last step is to consider the expression $e^{i\beta_1 \sigma_u^x} = k + il \sigma_u^x$. Each combination of terms is matched with its contribution to trace. If the first expression $e^{i\beta_1 \sigma_u^x}$ includes $il\sigma_u^x$, it will flip the sign of $V_1, V_4, W_3, W_4$. If the second expression $e^{i\beta_1 \sigma_u^x}$ includes $il\sigma_u^x$, it will flip the sign of $V_1, V_2, W_2, W_3$. 

\begin{enumerate}
    \item $(k, k) \to \Tr[(k)(k)(V_1 + V_2 + V_3 + V_4)(W_1 + W_2 + W_3 + W_4)^{D-1}\sigma_u^z\sigma_u^y ]$
    \item $(k, l) \to \Tr[(k)(il)(-V_1 - V_2 + V_3 + V_4)(W_1 - W_2 - W_3 + W_4)^{D-1}\sigma_u^z\sigma_u^y ]$
    \item $(l, k) \to \Tr[(il)(k)(-V_1 + V_2 + V_3 - V_4)(W_1 + W_2 - W_3 - W_4)^{D-1}\sigma_u^z\sigma_u^y]$
    \item $(l, l) \to \Tr[(il)(il)(V_1 - V_2 + V_3 - V_4)(W_1 - W_2 + W_3 - W_4)^{D-1}\sigma_u^z\sigma_u^y]$
\end{enumerate}

The $V$ and $W$ sums can then be simplified. Any expressions using $e^{ia \sigma^x}$ will become $e^{ia}$ after taking the trace. Here are the $V$ sums:
\begin{align*}
V_1 + V_2 + V_3 + V_4 
&= -impqe^{-2i\beta_1 } + -inp^2 + -impqe^{2i\beta_1 } + inq^2
= -impq(2r) - in(p^2 - q^2) 
= -imrz - iny   
\\
- V_1 - V_2 + V_3 + V_4 
&= - -impqe^{-2i\beta_1 } - -inp^2 + -impqe^{2i\beta_1 } + inq^2
= -impq(2it) +in(p^2 + q^2) 
= mtz + in
\\ 
-V_1 + V_2 + V_3 - V_4 
&= - -impqe^{-2i\beta_1 } + -inp^2 + -impqe^{2i\beta_1 } - inq^2 
= -impq(2it) - in(p^2 + q^2) 
= mtz - in
\\
V_1 - V_2 + V_3 - V_4 
&= -impqe^{-2i\beta_1 } - -inp^2 + -impqe^{2i\beta_1 } - inq^2 
= -impq(2r) +in(p^2 - q^2) 
= -imrz +iny
\end{align*}

Here are the $W$ sums:
\begin{align*}
W_1 + W_2 + W_3 + W_4 
&= mp^2 + -npqe^{2i \beta_1} + -mq^2 + -npqe^{-2i \beta_1}
& &= m(p^2 - q^2) - npq(2r) 
& &= my - nrz
\\
W_1 - W_2 - W_3 + W_4 
&= mp^2 - -npqe^{2i \beta_1} - -mq^2 + -npqe^{-2i \beta_1}
& &= m(p^2 + q^2) + npq(2it) 
& &= m + intz 
\\
W_1 + W_2 - W_3 - W_4 
&= mp^2 + -npqe^{2i \beta_1} - -mq^2 - -npqe^{-2i \beta_1}
& &= m(p^2 + q^2) -npq(2it) 
& &= m - intz
\\
W_1 - W_2 + W_3 - W_4 
&= mp^2 - -npqe^{2i \beta_1} + -mq^2 - -npqe^{-2i \beta_1}
& &= m(p^2 - q^2) + npq(2r) 
& &= my + nrz
\end{align*}

We sum all terms. Notice that any $V_i W_j$ will include an odd number of $\sigma_u^z$, which combined with $\sigma_u^y$ will add a factor of $-i$ to each term in the trace:
\begin{multline*}
B_1/cs =  (-i)
\Big(k^2 (-imrz - iny)(my - nrz)^{D-1} + ikl (mtz +in)(m + intz)^{D-1}
\\
+ ikl (mtz - in)(m - intz )^{D-1} -
l^2 (-imrz + iny)(my + nrz)^{D-1} \Big)
\end{multline*}

In a simplified form,
\begin{multline*}
   B_1/cs =0.5 \Big( (1+r)(-mrz - ny)(my - nrz)^{D-1} +
t(mtz + in)(m + intz)^{D-1}
\\
+ t (mtz - in)(m - intz)^{D-1} +
(1-r) (mrz - ny)(my + nrz)^{D-1} \Big)
\end{multline*}

\subsubsection{Effect of the crossover terms}
\label{sub:b1crossover}
Now we consider the crossover terms.
For edges that introduced a new node $k$, the associated $\sigma_z^k$ must be cancelled. Since the graph has girth above 5, any neighbor $k \notin S$ to $\ell \in S$ has a unique neighbor $\ell$. So, the above expression is modified by 
$$
\Tr[(p - iq \sigma_j^z \sigma_k^z) \cdots (p + iq \sigma_j^z \sigma_k^z)] = p^2 \Tr[\cdots] + q^2 [\sigma_j^z \cdots \sigma_j^z]
$$

For $\ell =v$, the only term that does not commute with $\sigma_v^z$ is $e^{2 i \beta_1 \sigma_{v}^x} m \to e^{-2 i \beta_1 \sigma_{v}^x} m $. This swaps $V_1$ and $V_3$, which only adjusts $mtz \to -mtz$ in the 2nd and 3rd term of the above expression. Thus, $mtz \to (p^2 - q^2) mtz = mtyz$. All other terms are $T \to (p^2 + q^2)T = T$. This process happens $D-1$ times (once for each neighbor $x_j\ne u$ of $v$).
\begin{multline*}
   B_1/cs =0.5 \Big( (1+r)(-mrz - ny)(my - nrz)^{D-1} +
t(mty^{D-1}z + in)(m + intz)^{D-1} +
\\
t (mty^{D-1}z - in)(m - intz)^{D-1} +
(1-r) (mrz - ny)(my + nrz)^{D-1} \Big)
\end{multline*}

Consider $\ell= w_i$ for $i \in \{1, \cdots, D-1\}$. The only term that does not commute with $\sigma_{w_i}^z$ is $e^{2 i \beta_1 \sigma_{w_i}^x} \to e^{-2 i \beta_1 \sigma_{w_i}^x}$. 
This swaps $W_2$ and $W_4$, which converts $(m \pm intz)$ to $(m \mp intz)$. 
Thus, $(m \pm intz) \to (p^2 (m \pm intz) + q^2 (m \mp intz) = (m \pm intyz)$. All other terms are $T \to (p^2 + q^2)T = T$. This process happens $D-1$ times (once for each neighbor $x_j \ne u$ of $w_i$). There are $D-1$ nodes of the form $w_i$.
\begin{align*}
   B_1/cs &=0.5 \big( (1+r)(-mrz - ny)(my - nrz)^{D-1} +
t(mty^{D-1}z + in)(m + inty^{D-1}z)^{D-1}
\\
&+ t (mty^{D-1}z - in)(m - inty^{D-1}z)^{D-1} +
(1-r) (mrz - ny)(my + nrz)^{D-1} \big)
\\
\\ &:= 0.5 \alpha(\gamma_1, \beta_1, \gamma_2) 
\end{align*}

\subsection{Calculating $E$}
\subsubsection{Evaluating exponentials}
Most terms of $C$ commute with $\sigma_u^y \sigma_v^y$. The ones that do not correspond to the other $D-1$ neighbors of $u$ and $v$. 
$$
E/s^2 = \Tr[\rho_0 e^{i\gamma_1 C} e^{i\beta_1 J}
\big( \prod_{i=1}^{D-1} e^{-i \gamma_2 \sigma_u^z \sigma_{w_i}^z} \big) 
\big( \prod_{j=1}^{D-1} e^{-i \gamma_2 \sigma_v^z \sigma_{\chi_j}^z} \big) 
 \sigma_u^y \sigma_v^y
e^{-i\beta_1 J} e^{-i\gamma_1 C} ]
$$

The only terms of $J$ that do not commute with the inner expression involve $u, v, w_1, ..., w_{D-1}, \chi_1, ..., \chi_{D-1}$.
\begin{multline*}
E/s^2 = \text{Tr}\Bigg[\rho_0 e^{i\gamma_1 C} 
e^{i\beta_1 \sigma_u^x}
e^{i\beta_1 \sigma_v^x}
\big( \prod_{i=1}^{D-1} 
    e^{i\beta_1 \sigma_{w_i}^x} 
    e^{-i \gamma_2 \sigma_u^z \sigma_{w_i}^z} 
    e^{-i\beta_1 \sigma_{w_i}^x}\big) 
\\
\big( \prod_{j=1}^{D-1} 
    e^{i\beta_1 \sigma_{\chi_j}^x} 
    e^{-i \gamma_2 \sigma_v^z \sigma_{\chi_j}^z} 
        e^{-i\beta_1 \sigma_{\chi_j}^x}\big) 
e^{i\beta_1 \sigma_u^x}
e^{i\beta_1 \sigma_v^x}
 \sigma_u^y \sigma_v^y
e^{-i\gamma_1 C} \Bigg]
\end{multline*}

The terms within the parentheses can be simplified.
$$
e^{i \beta_1 \sigma_{w_i}^x}  e^{-i \gamma_2 \sigma_u^z \sigma_{w_i}^z} e^{-i\beta_1 \sigma_{w_i}^x}
 = e^{i \beta_1 \sigma_{w_i}^x} (m - in \sigma_u^z \sigma_{w_i}^z) e^{-i \beta_1 \sigma_{w_i}^x}
 = m - in e^{2 i \beta_1 \sigma_{w_i}^x}\sigma_u^z \sigma_{w_i}^z
$$
$$
E/s^2 = \Tr[\rho_0 e^{i\gamma_1 C} 
e^{i\beta_1 \sigma_u^x}
e^{i\beta_1 \sigma_v^x}
\big( \prod_{i=1}^{D-1} 
m - in e^{2 i \beta_1 \sigma_{w_i}^x}\sigma_u^z \sigma_{w_i}^z
\big) 
\big( \prod_{j=1}^{D-1} 
m - in e^{2 i \beta_1 \sigma_{\chi_j}^x}\sigma_v^z \sigma_{\chi_j}^z
\big) 
e^{i\beta_1 \sigma_u^x}
e^{i\beta_1 \sigma_v^x}
 \sigma_u^y \sigma_v^y
e^{-i\gamma_1 C} ]
$$

\subsubsection{Breaking apart into terms}

 Terms of $e^{i\gamma_1 C}$ that connect $j$ and $k$ have the form $p - iq \sigma_j^z \sigma_k^z$. If a term of $C$ does not commute, it must involve an edge connected to a node in $R = \{u, v, w_1, ..., w_{D-1}, \chi_1, ..., \chi_{D-1}\}$. For now, consider only edges where $j, k \in R$. We consider the ``crossover terms" in Appendix \ref{sub:ecrossover}. Because of $\rho_0$, the trace is nonzero only for terms that are proportional to $I$ or $\sigma_n^x$ for all nodes $n$.
\begin{multline*}
    E/s^2 =  
\text{Tr}\Bigg[ \rho_0 
(p - iq \sigma_u^z \sigma_v^z)
\big( \prod_{i=1}^{D-1} (p - iq \sigma_u^z \sigma_{w_i}^z) \big)
\big( \prod_{j=1}^{D-1} (p - iq \sigma_v^z \sigma_{\chi_j}^z) \big)
\\ e^{i\beta_1 \sigma_u^x}
e^{i\beta_1 \sigma_v^x}
\big( \prod_{i=1}^{D-1} 
m - in e^{2 i \beta_1 \sigma_{w_i}^x}\sigma_u^z \sigma_{w_i}^z
\big) 
\big( \prod_{j=1}^{D-1} 
m - in e^{2 i \beta_1 \sigma_{\chi_j}^x}\sigma_v^z \sigma_{\chi_j}^z
\big) 
e^{i\beta_1 \sigma_u^x}
e^{i\beta_1 \sigma_v^x}
\\ (p + iq \sigma_u^z \sigma_v^z)
\big( \prod_{i=1}^{D-1} (p - iq \sigma_u^z \sigma_{w_i}^z) \big)
\big( \prod_{j=1}^{D-1} (p - iq \sigma_v^z \sigma_{\chi_j}^z) \big)
\sigma_u^y \sigma_v^y
\Bigg]
\end{multline*}

Any nonzero trace terms including $\sigma_z$ or $\sigma_y$ must be converted to $I$ or $\sigma_x$. Of the expressions that include $\sigma_{w_i}^z$, only half include an even number. Only these corresponding terms will contribute to trace. Note that each of these terms includes an even number of $\sigma_u^z$ which cancel. Sometimes the exponential sign flips when commuting with $\sigma^z$. This entire process also holds for $\sigma_{\chi_j}^z$; let's call those terms $X_1, X_2, X_3, X_4$.
$$
(p - iq \sigma_u^z \sigma_{w_i}^z)
 \cdots (m - in e^{2 i \beta_1 \sigma_{w_i}^x}\sigma_u^z \sigma_{w_i}^z)
\cdots (p - iq \sigma_u^z \sigma_{w_i}^z)
$$

\begin{enumerate}
    \item  $(p, m, p) \to W_1 = mp^2$
    \item $(p, - in e^{2 i \beta_1 \sigma_{w_i}^x}\sigma_u^z \sigma_{w_i}^z, - iq \sigma_u^z \sigma_{w_i}^z) \to W_2 = -npqe^{2 i \beta_1 \sigma_{w_i}^x}$
    \item $(- iq \sigma_u^z \sigma_{w_i}^z, m, - iq \sigma_u^z \sigma_{w_i}^z) \to W_3 = -mq^2$
    \item  $( - iq \sigma_u^z \sigma_{w_i}^z, - in e^{2 i \beta_1 \sigma_{w_i}^x}\sigma_u^z \sigma_{w_i}^z, p) \to W_4 = -npq e^{-2 i \beta_1 \sigma_{w_i}^x}$
\end{enumerate}

Consider the $\sigma_u^y \sigma_v^y$ term. This can only be cancelled with $(p \pm iq \sigma_u^z \sigma_v^z)$, since other terms include additional $\sigma_{w_i}^z$ or $\sigma_{\chi_j}^z$ which cannot be cancelled out alone. These are the only allowed terms from $(p - iq \sigma_u^z \sigma_v^z) \cdots (p + iq \sigma_u^z \sigma_v^z) \sigma_u^y \sigma_v^y$:
\begin{enumerate}
    \item $(p)\cdots(iq\sigma_u^z \sigma_v^z) \sigma_u^y \sigma_v^y \to -ipq$
    \item $(-iq\sigma_u^z \sigma_v^z)\cdots(p)\sigma_u^y \sigma_v^y \to ipq$, where all $e^{i\beta_1 \sigma_u^x}e^{i\beta_1 \sigma_v^x} \to e^{-i\beta_1 \sigma_u^x}e^{-i\beta_1 \sigma_v^x}$
\end{enumerate}
Expanding the mentioned exponentials and summing the two terms simplifies to 
$$
2pqkl(k^2 - l^2 \sigma_u^x \sigma_v^x) \cdots (\sigma_u^x + \sigma_v^x)
$$
plus the terms reversed:
$$
2pqkl (\sigma_u^x + \sigma_v^x)\cdots(k^2 - l^2 \sigma_u^x \sigma_v^x)
$$
Each combination of terms is matched with its contribution to trace. Note that if the first expression includes $\sigma_u^x$, it will flip the sign of $W_3, W_4$. If the second expression includes $\sigma_u^x$, it will flip the sign of $W_2, W_3$. The same is true for $\sigma_v^x$ and $X$ terms.
\begin{enumerate}
    \item $(k^2, \sigma_u^x) \to \Tr[2pqkl(k^2)(W_1 - W_2 - W_3 + W_4)^{D-1} (X_1 + X_2 + X_3 + X_4)^{D-1}]$
    \item $(k^2, \sigma_v^x) \to \Tr[2pqkl(k^2)(W_1 + W_2 + W_3 + W_4)^{D-1} (X_1 - X_2 - X_3 + X_4)^{D-1}]$
    \item $(\sigma_u^x, k^2) \to \Tr[2pqkl(k^2)(W_1 + W_2 - W_3 - W_4)^{D-1} (X_1 + X_2 + X_3 + X_4)^{D-1}]$
    \item $(\sigma_v^x, k^2) \to \Tr[2pqkl(k^2)(W_1 + W_2 + W_3 + W_4)^{D-1} (X_1 + X_2 - X_3 - X_4)^{D-1}]$
    \item $(-l^2\sigma_u^x \sigma_v^x, \sigma_u^x) \to \Tr[2pqkl(-l^2)(W_1 - W_2 + W_3 - W_4)^{D-1} (X_1 + X_2 - X_3 - X_4)^{D-1}]$
    \item $(-l^2\sigma_u^x \sigma_v^x, \sigma_v^x) \to \Tr[2pqkl(-l^2)(W_1 + W_2 - W_3 - W_4)^{D-1} (X_1 - X_2 + X_3 - X_4)^{D-1}]$    
    \item $(\sigma_u^x, -l^2\sigma_u^x \sigma_v^x) \to \Tr[2pqkl(-l^2)(W_1 - W_2 + W_3 - W_4)^{D-1} (X_1 - X_2 - X_3 + X_4)^{D-1}]$
    \item $( \sigma_v^x, -l^2\sigma_u^x \sigma_v^x) \to \Tr[2pqkl(-l^2)(W_1 - W_2 - W_3 + W_4)^{D-1} (X_1 - X_2 + X_3 - X_4)^{D-1}]$   
\end{enumerate}

After taking the trace, each of $W_\ell = X_\ell$. Summing all 8 combinations:
\begin{align*}
    E/(s^2 2pqkl) &= 2k^2 (W_1 + W_2 + W_3 + W_4)^{D-1} \big( (W_1 - W_2 - W_3 + W_4)^{D-1} + (W_1 + W_2 - W_3 - W_4)^{D-1} \big)
    \\
    &- 2l^2 (W_1 - W_2 + W_3 - W_4)^{D-1} \big( (W_1 - W_2 - W_3 + W_4)^{D-1} + (W_1 + W_2 - W_3 - W_4)^{D-1} \big)
    \\
    \\
    &= 2k^2 (my - nrz)^{D-1} \big( (m + intz)^{D-1} + (m - intz)^{D-1} \big) 
    \\
    &- 2l^2 (my + nrz)^{D-1} \big( (m + intz)^{D-1} + (m - intz)^{D-1} \big)
    \\
    \\
    E/(s^2 tz) &= \big(k^2 (my - nrz)^{D-1} - l^2 (my + nrz)^{D-1}\big)\big( (m + intz)^{D-1} + (m - intz)^{D-1} \big) 
    \\
    &= 0.5\big((1+r) (my - nrz)^{D-1} - (1-r) (my + nrz)^{D-1}\big)\big( (m + intz)^{D-1} + (m - intz)^{D-1} \big)
\end{align*}

\subsubsection{Effect of the crossover terms}
\label{sub:ecrossover}

Now we consider the crossover terms. For edges that introduced a new node $k$, the associated $\sigma_z^k$ must be cancelled. Since the graph has girth above 5, any neighbor $k \notin R$ to $\ell \in R$ has a unique neighbor $\ell$. So, the above expression is modified by 
$$
\Tr[(p - iq \sigma_j^z \sigma_k^z) \cdots (p + iq \sigma_j^z \sigma_k^z)] = p^2 \Tr[\cdots] + q^2 [\sigma_j^z \cdots \sigma_j^z]
$$

Consider $\ell = w_i$ for $i \in \{1, \cdots, D-1\}$. The only term that does not commute with $\sigma_{w_i}^z$ is $e^{2 i \beta_1 \sigma_{w_i}^x} \to e^{-2 i \beta_1 \sigma_{w_i}^x}$. 
This swaps $W_2$ and $W_4$, which converts $(m \pm intz)$ to $(m \mp intz)$. 
Thus, $(m \pm intz) \to (p^2 (m \pm intz) + q^2 (m \mp intz) = (m \pm intyz)$. All other terms are $T \to (p^2 + q^2)T = T$. This process happens $D-1$ times (once for each neighbor $x_j \ne u$ of $w_i$). There are $D-1$ nodes of the form $w_i$. The same process repeats for $\ell = \chi_j$.
\begin{align*}
    E/(s^2 tz) &= 0.5\big((1+r) (my - nrz)^{D-1} - (1-r) (my + nrz)^{D-1}\big)
    \\
    &\times \big( (m + inty^{D-1}z)^{D-1} + (m - inty^{D-1}z)^{D-1} \big) 
    \\
    \\
    &:= 0.5\kappa(\gamma_1, \beta_1, \gamma_2)
\end{align*}

\subsection{Full expression}
Putting all expressions together:
$$
 f_2(\gamma_1, \beta_1, \gamma_2, \beta_2) = \frac{1}{2} - \frac{1}{2} (A + 2B_1 + E) 
= \frac{1}{2} + c^2 rt y^{D-1} z - \frac{cs}{2} \alpha(\gamma_1, \beta_1, \gamma_2) - \frac{s^2 tz}{4} \kappa(\gamma_1, \beta_1, \gamma_2)
$$
This completes the proof.

\clearpage

\section{Correctness tests of the $p=2$ QAOA formula}
\label{sec:correctness_qaoa2formula}
To verify this result, we apply restrictions to $D$ and the input parameters to recover known expressions.

\subsection{Simplification to $p=1$ formula}
First suppose $\gamma_1 = \beta_1 = 0$. Then $r = y = 1$ and $t = z = 0$. We verify that $f_2(0, 0, \gamma_2, \beta_2) = f_1(\gamma_2, \beta_2)$. We simplify $\alpha$:
\begin{align*}
\alpha(0, 0, \gamma_2) &= (1+r)(-mrz - ny)(my - nrz)^{D-1} +(1-r) (mrz - ny)(my + nrz)^{D-1}
\\ &+ t \big ( (mty^{D-1}z + in)(m + inty^{D-1}z)^{D-1} + (mty^{D-1}z - in)(m - inty^{D-1}z)^{D-1} \big )
\\
&= 2(-n)(m)^{D-1} = -2nm^{D-1}
\end{align*}

Thus, $f_2(0, 0, \gamma_2, \beta_2)$ matches the $p=1$ formula given in Corollary 1 of \cite{Wang2018}:
\begin{align*}
 f_2(0, 0, \gamma_2, \beta_2)  
&= \frac{1}{2} + c^2 rt y^{D-1} z - \frac{cs}{2} \alpha(0, 0, \gamma_2) - \frac{s^2 tz}{4} \kappa(0, 0, \gamma_2)
\\
&= \frac{1}{2} + 0 + csnm^{D-1} + 0 = \frac{1}{2} + \sin(2\beta_2)\cos(2\beta_2) \sin(\gamma_2)\cos^{D-1}(\gamma_2)
\end{align*}

Now, instead suppose $\gamma_2 = \beta_2 = 0$. Then $c = m = 1$ and $s = n = 0$. Then $f_2( \gamma_1, \beta_1, 0, 0) = f_1(\gamma_1, \beta_1)$:
$$
f_2( \gamma_1, \beta_1, 0, 0) = \frac{1}{2} + rty^{D-1}z + 0 + 0 = \frac{1}{2} + \sin(2\beta_1)\cos(2\beta_1)\sin(\gamma_1)\cos^{D-1}(\gamma_1)
$$

\subsection{Simplification to the ring ($D=2$)}
Consider the case when $D = 2$. We simplify $\alpha$:
\begin{align*}
\alpha(\gamma_1, \beta_1, \gamma_2) &= (1+r)(-mrz - ny)(my - nrz)^{D-1} +(1-r) (mrz - ny)(my + nrz)^{D-1}
\\ &+ t \big ( (mty^{D-1}z + in)(m + inty^{D-1}z)^{D-1} + (mty^{D-1}z - in)(m - inty^{D-1}z)^{D-1} \big )
\\
\\
&= (1+r)(-mrz - ny)(my - nrz) +(1-r) (mrz - ny)(my + nrz)
\\ &+ t \big ( (mtyz + in)(m + intyz) + (mtyz - in)(m - intyz) \big )
\\
\\
&= 2(-mny^2 + mnr^2 z^2) + 2r(-m^2 ryz + n^2 ryz)
\\ &+ t \big ( 2m^2 tyz - 2n^2 tyz  \big )
\\
\\
&= 2mn(r^2 z^2 - y^2) + 2yz(t^2-r^2)(m^2 - n^2)
\end{align*}

We also simplify $\kappa$:
\begin{align*}
\kappa(\gamma_1, \beta_1, \gamma_2) &= \big((1+r)(my - nrz)^{D-1} - (1-r)(my + nrz)^{D-1}\big)\big( (m + inty^{D-1}z)^{D-1} + (m - inty^{D-1}z)^{D-1} \big)
\\
&= \big((1+r)(my - nrz) - (1-r)(my + nrz)\big)\big( (m + intyz) + (m - intyz) \big)
\\
&= \big(2rmy - 2nrz \big)(2m) = 4mr(my - nz)
\end{align*}

Then we simplify $f_2$:
\begin{align*}
 f_2(\gamma_1, \beta_1, \gamma_2, \beta_2)
&= \frac{1}{2} + c^2 rt y^{D-1} z - \frac{cs}{2} \alpha - \frac{s^2 tz}{4} \kappa
\\
&= \frac{1}{2} + c^2 rtyz - csmn(r^2 z^2 - y^2) - csyz(t^2 - r^2)(m^2 - n^2) - s^2 tz mr(my - nz)
\end{align*}

Using SymPy \cite{sympy}, we verify that this expression is symbolically equal to the expression described in Section IV and Appendix C of \cite{Wang2018}. The code is \href{https://nbviewer.jupyter.org/github/marwahaha/qaoa-local-competitors/blob/master/2-step-comparison.ipynb}{available online}.
\clearpage

\section{Analysis of the 2-step threshold algorithm}
\label{sec:threshold2_analysis}
This follows the analysis given in \cite{hastings2019classical}. Consider a subgraph around an edge $E_{ij}$ of a $(n+1)$-regular graph. Let $P_n(k) = 2^{-n} { n \choose k}$ be the binomial probability mass function. Define the threshold functions $q_i(k)$: -1 if $k \ge \tau_i$ and 1 otherwise for $i \in \{1,2\}$. Let the graph have girth $> 5$; that is, no neighbors of $i$ are adjacent to $j$ nor neighbors of $j$, and vice versa. We calculate the expectation $\langle Z_{i2} Z_{j2} \rangle$, which is $-2$ times the improvement over random assignment.

\subsection{$Z_i, Z_j$ initially equal}

Suppose $Z_{i0} = 1, Z_{j0} = 1$, where $k/n$ other neighbors of $i$ initially agree with $Z_i$, and $u/n$ other neighbors of $j$ agree with $Z_j$.
Then, $Z_{i1,k} = q_1(k+1)$ and $Z_{j1,u} = q_1(u+1)$  (which could be $\pm 1$ depending on the threshold). The average spin of a neighbor of $i$ (or $j$) is $Z_{1,+} = (+1)\sum_{m=0}^n P_n(m) q_1(m+1)$ if they initially agreed with $i$ (or $j$), or $Z_{1,-} = (-1) \sum_{m=0}^n P_n(m) q_1(m)$ if they didn't agree initially.
\newline

Since the actual spin can only be $\pm 1$, let's define the following:
\begin{align*}
    p_+ &= P(Z_{1,+} = 1) = (1 + Z_{1,+}) / 2 
    & p_- &= P(Z_{1,-} = 1) = (1 + Z_{1,-}) / 2
    \\
    P1^+_k(l) &=  p_+^l (1-p_+)^{k-l} {k \choose l}
    & P1^-_k(l) &= p_-^l (1-p_-)^{k-l} {k \choose l}
    \\
    H(k,l,r) &= P1^+_k(l)P1^-_{n-k}(r)
\end{align*}
$H(k,l,r)$ describes the chance of $l+r$ neighbors agreeing with your initial value, given $k/n$ initially agreeing with your initial value. 

\begin{enumerate}
\item Suppose $q_1(k+1) = 1$ and $q_1(u+1) = 1$ (both under threshold). Then:
$$
Z_{i2,k} = \sum_{l=0}^k \sum_{r=0}^{n-k} H(k,l,r) q_2(l+r+1)
$$
$$
Z_{j2,u} = \sum_{l=0}^u \sum_{r=0}^{n-u} H(u,l,r)q_2(l+r+1)
$$
\item Suppose instead $q_1(k+1) = -1$ and $q_1(u+1) = -1$. Then, there are $l+r$ neighbors with spin $+1$, so $n-l-r$ neighbors with spin matching $Z_{i1}$, plus $1$ because $Z_{i1} = Z_{j1}$.
$$
Z_{i2,k} = (-1) \sum_{l=0}^k \sum_{r=0}^{n-k} H(k,l,r) q_2(n-l-r+1)
$$
$$
Z_{j2,u} = (-1) \sum_{l=0}^u \sum_{r=0}^{n-u} H(u,l,r)q_2(n-l-r+1)
$$
\item Suppose instead $q_1(k+1) = 1$ and $q_1(u+1) = -1$. This is similar to before, but minus $1$ since $Z_{i1} \ne Z_{j1}$. 
$$
Z_{i2,k} = \sum_{l=0}^k \sum_{r=0}^{n-k} H(k,l,r) q_2(l+r)
$$
$$
Z_{j2,u} = (-1) \sum_{l=0}^u \sum_{r=0}^{n-u} H(u,l,r)q_2(n-l-r)
$$
\item The last case is the same as the previous one after exchanging $Z_i$ with $Z_j$.
\end{enumerate}

We can define $Q$ to handle the different cases:
$$
Q(l,r, A, B) = q_2\big( (n-l-r)(1-A)/2 + (l+r)(1+A)/2 + (1 + B)/2)\big)
$$

$$
Z_{i2,k} = Z_{i1,k} \sum_{l=0}^k \sum_{r=0}^{n-k} H(k,l,r) Q(l,r,Z_{i1,k}Z_{i0}, Z_{i1,k}Z_{j1,u})
$$

$$
Z_{j2,u} = Z_{j1,u} \sum_{l=0}^u \sum_{r=0}^{n-u} H(u,l,r) Q(l,r,Z_{j1,u}Z_{j0}, Z_{j1,u}Z_{i1,k})
$$

$Q$ will use $l+r$ if $A = 1$ (i.e. under threshold), otherwise $n-l-r$. It will add one iff B = 1, i.e. $Z_{i1} = Z_{j1}$.
\newline

In sum:

$$
\langle Z_{i2} Z_{j2} \rangle = \sum_{k=0}^n \sum_{u=0}^n P_n(k) P_n(u) Z_{i2,k} Z_{j2,u}
$$

This equation is symmetric; if $Z_{i0} = Z_{j0} = -1$, the result is the same.

\subsection{$Z_i, Z_j$ initially unequal}
Suppose instead that $Z_{j0} = -1$, and still $u/n$ of its other neighbors agree with $j$. Then, $Z_{i1,k} = q_1(k)$, and $Z_{j1,u} = -q_1(u)$. So the above expressions for $Z_{i2,k}, Z_{j2,u}$, and $\langle Z_{i2} Z_{j2} \rangle$ still hold with new values of $Z_1$. This equation is symmetric; if $Z_{i0} = -1$ and $Z_{j0} = 1$, the result is the same.

\subsection{All together}
It is equally likely for $i$ and $j$ to initially agree or disagree. So the spin correlation is as follows:

\begin{align*}
    \langle Z_{i2} Z_{j2} \rangle = \sum_{k=0}^n \sum_{u=0}^n P_n(k) P_n(u)  0.5 \big(Z_{i2,k} Z_{j2,u}&\Big\rvert_{Z_{i1,k} = q_1(k+1),Z_{j1,u} = q_1(u+1),Z_{i0}=Z_{j0} = 1} \\
    + Z_{i2,k} Z_{j2,u}&\Big\rvert_{Z_{i1,k} = q_1(k),Z_{j1,u} = -q_1(u), Z_{i0}=1, Z_{j0} = -1}\big)
\end{align*}

\end{document}